\pdfoutput=0
\documentclass{aa}

\usepackage{epsfig,graphics,color}
\usepackage {txfonts }
\usepackage {natbib  }

\begin{document}

%
%

\title {Site testing for submillimetre astronomy at Dome C, Antarctica
       }


\author{ P. Tremblin  \inst{1}\and 
V. Minier        \inst{1}\and   
N. Schneider     \inst{1}\and       
G. Al. Durand    \inst{1}\and
M. C. B. Ashley  \inst{2}\and 
J. S. Lawrence   \inst{2}\and 
D. M. Luong-Van  \inst{2}\and 
J. W. V. Storey  \inst{2}\and
G. An. Durand    \inst{3}\and
Y. Reinert       \inst{3}\and
C. Veyssiere     \inst{3}\and
C. Walter        \inst{3}\and
P. Ade           \inst{4}\and
P. G. Calisse    \inst{4}\and
Z. Challita      \inst{5,6}\and 
E. Fossat        \inst{6}  \and
L. Sabbatini     \inst{5,7}\and
A. Pellegrini    \inst{8}  \and   
P. Ricaud        \inst{9}  \and   
J. Urban         \inst{10}  
       }

\institute{Laboratoire AIM Paris-Saclay (CEA/Irfu - Uni. Paris Diderot - CNRS/INSU), 
           Centre d'\'etudes de Saclay,  91191 Gif-Sur-Yvette, France    
           \and 
           University of New South Wales, 2052 Sydney, Australia 
           \and  
           Service d'ing\'enierie des syst\`emes, CEA/Irfu, Centre d'\'etudes de
           Saclay, 91191 Gif-Sur-Yvette, France
           \and 
           School of Physics \& Astronomy, Cardiff University, 5 The Parade, Cardiff, CF24 3AA, UK
           \and
           Concordia Station, Dome C, Antarctica
           \and
           Laboratoire Fizeau (Obs. Côte d'Azur - Uni. Nice Sophia Antipolis - CNRS/INSU), 
           Parc Valrose, 06108 Nice, France
           \and 
           Departement of Physics, University of Roma Tre, Italy
           \and 
           Programma Nazionale Ricerche in Antartide, ENEA, Rome Italy
           \and 
           Laboratoire d'A\'erologie, UMR 5560 CNRS-Universit\'e Paul-Sabatier, 
           31400 Toulouse, France 
           \and 
           Chalmers University of Technology, Department of Earth and Space Sciences, 
           41296 G\"oteborg, Sweden    
           }

\date{
}
\offprints{P. Tremblin}
\mail{pascal.tremblin@cea.fr, vincent.minier@cea.fr}

\titlerunning{Site testing Dome C}
\authorrunning{P. Tremblin et al.}

\abstract{}
{
Over the past few years a major effort has been put into the exploration 
of potential sites for the deployment of submillimetre astronomical facilities.  
Amongst the most important sites are Dome C and Dome A on the Antarctic
Plateau, and the Chajnantor area in Chile. In this context, we report on
measurements of the sky opacity at 200 $\mu$m over a period of three
years at the French-Italian station, Concordia, at Dome C, Antarctica.  We also present some solutions to the challenges of operating in the harsh polar environment.
}
{ 
The 200-$\mu$m atmospheric opacity was measured with a tipper. The forward atmospheric model MOLIERE (Microwave
Observation LIne Estimation and REtrieval) was used to calculate the
atmospheric transmission and to evaluate the precipitable water vapour content (PWV)
from the observed sky opacity.   These results have been compared with satellite
measurements from the Infrared Atmospheric Sounding Interferometer
(IASI) on Metop-A, with balloon humidity sondes and with results obtained by a ground-based
microwave radiometer (HAMSTRAD).  In addition, a series of experiments has been designed to study 
frost formation on surfaces, and the temporal and spatial evolution of thermal gradients in the low
atmosphere.
}
{
Dome C offers exceptional conditions in terms of absolute atmospheric
transmission and stability for submillimetre astronomy.   Over
the austral winter the
PWV exhibits long periods during which it is stable and at a very low level (0.1 to 0.3 mm). Higher values (0.2 to 0.8 mm) of PWV are observed
during the short summer period. Based on observations over three years, a
transmission of around 50\% at 350 $\mu$m is achieved for 75\% of the
time. The 200-$\mu$m window opens with a typical transmission of
10\% to 15\% for 25\% of the time.
}
{
Dome C is one of the best  accessible sites on Earth for
submillimetre astronomy. Observations at 350 or 450 $\mu$m are
possible all year round, and the 200-$\mu$m window opens long enough and with
a sufficient transparency to be useful. Although the polar environment severely constrains
hardware design, a permanent observatory with appropriate technical capabilities is
feasible. Because of the very good astronomical conditions, high angular resolution and
time series (multi-year) observations at Dome C with a medium size single dish
telescope would enable unique studies to be conducted, some of which are not otherwise feasible even from space.
}


\titlerunning {Site Testing at Dome C}
\authorrunning{Tremblin et al.       }

\maketitle

%
%

\section{Introduction}
Submillimetre (submm) astronomy is one of the most important techniques for the study of the cold Universe and for unveiling the birth
and early evolution of planets, stars, and galaxies. Submm
continuum observations are particularly powerful for the measurement of
luminosities, temperatures, and masses of cool, dusty objects
because dust enshrouded star-forming regions emit the bulk of their
energy between 60 and 500 $\mu$m. The submm range of the spectrum
(or THz regime in frequency) is also rich in the atomic and molecular lines
that offer the only means of studying the kinematic structure of the
interstellar medium of galaxies. 
Observations at these
wavelengths with medium to large telescopes should lead
to breakthroughs in the study of star-formation at all scales, and to a better
understanding of its history back to the early Universe---thus leading to a
better understanding of galaxy evolution. Asteroids, debris disks,
planet formation, dust origin in evolved stars, interstellar dust and
polarisation of dust in the Universe are also potential science
drivers for submm astronomy \citep[see][for details]{Minier2009}. 

\subsection{Dome C, a potential site}

A major obstacle to ground-based observations in the submm range (and specifically at wavelengths shorter than 500 $\mu$m) is the atmosphere. 
This part of the electromagnetic spectrum is normally the preserve of space telescopes such as the Herschel space observatory \citep{Pilbratt2010}
although large submm facilities such as ALMA will be able to operate down to 420 $\mu$m and possibly below in
the future  \citep{Hills2010}. However, submm observations in the 200-$\mu$m window with ground-based instruments will always 
require exceptional conditions \citep[see][]{Marrone2005,Oberst2006}. A very good, high-altitude 
site like Chajnantor is usable for less than $25\%$ of the time in winter (i.e. at least 40\% transmission) 
in the 350/450 $\mu$m windows, and probably for less than $5\%$ of the time at 200 $\mu$m \citep{Peterson2003}.  \citet{Matsushita1999} reported a more optimistic estimate, operational experience with APEX, ALMA, and other telescopes will tell.

While ground-based observations are limited by the atmosphere, space telescopes remain limited in size and 
thus can offer only modest angular resolution: from $\sim$6$''$ to 37$''$ at 70 to 500 $\mu$m, respectively
for the Herschel space observatory, implying a fairly high extragalactic confusion limit and preventing
the study of individual protostars in all but the nearest star-forming clusters of our Galaxy \citep[e.g.][]{Schneider2010,Andre2010}. 
In this context, large ($>10$ m) single-dish telescopes operating at 200-450 $\mu$m, can provide a better angular resolution 
than Herschel, and wider-field mapping capabilities than ALMA.  This appears to satisfy a clear need in submm astronomy (e.g. Radford et al. 2008).
As a result, an intensive study of new sites is currently underway.

Although the average altitude of the Antarctic plateau, $\sim$3000m, is less than that of the Chajnantor plateau ($\sim$5000m), Antarctica
might nevertheless offer better conditions for submm astronomy because of its peculiar geography and climate. The very low atmospheric precipitable water vapour content (PWV) in Antarctic results from the low sun elevation, isolation by the circumpolar vortex , and the high optical reflectivity of ice combined with its high emissivity in the infrared---leading to intense radiative cooling of the ice at night. 
These phenomena make Antarctica the coldest continent on Earth. Snow precipitation is very low on the Antarctic 
plateau and low pressure fronts rarely penetrate into the inner plateau, instead staying
at the coastline. In fact, the inland part of the continent is a true
desert: across an area of 5 million km$^2$, snow precipitation is
about 5 cm annually, and often less than 2 cm at the highest regions (called
Domes). As a consequence, the average PWV above the Antarctic plateau is expected
to be lower than at Chajnantor  \citep[][and references
therein]{Minier2009}. 

These high plateau Antarctic sites (Domes A, C, F) are therefore
potential locations for astronomical facilities \citep[for Dome A see][]{Yang2010}. 
The relative merits of the sites have been presented in \citet{Saunders2009}. These studies follow the 
development of astronomy at the South Pole, where submm facilities have already been deployed
by US astronomers. The atmospheric transmission at 225 GHz has been  
measured there in 1992 with an NRAO radiometer \citep{Chamberlin2004}.The pathfinder for Antarctic submm-astronomy was AST/RO, a 1.7 m dish 
\citep{Stark2001} installed at the South Pole and operational from 1995. 
Recently, a 10m submm dish has been built at the South Pole 
\citep[South Pole Telescope SPT,][]{Carlstrom2011}. 

Dome C is the location of the French-Italian Concordia station,
accessible from both the Dumont-d'Urville and Mario Zucchelli coastal stations 
by either light plane or, for transporting heavy material,
motorised ground expeditions. Concordia station
hosts a crew of people during winter and thus allows
experiments all year round. Monitoring of the atmosphere characteristics and
qualification of the site for optical and near-infrared astronomy have
been conducted for many years by French, Italian, and Australian teams
\citep[e.g.][]{Aristidi2005,Lawrence2004}. However, 
little effort has so far been put into evaluation of the
quality of the atmosphere, the meteorological constraints, or the
specific advantages of Dome C for a potential submm/FIR telescope. Preliminary
meteorological studies and atmospheric transmission models (\citet{Schneider2009}
and Fig. \ref{moliere_fig}) suggest that Dome C might offer
atmospheric conditions that open the 200-$\mu$m window 
\citep{Valenziano1999, Tomasi2006, Minier2008} for a
significant amount of time. \citet{Calisse2004} undertook
measurements of the atmospheric opacity at 350 $\mu$m during a summer
campaign and found that it is comparable, but consistently better than, that at South Pole. However, no
direct assessment of the wintertime atmospheric transmission has so far been
performed.

The stability of the atmosphere is an equally important parameter when
comparing different sites, and Dome C may turn out to be far more
stable than the Chilean sites \citep{Minier2008}. Determining whether a telescope
facility at Dome C in Antarctica might be able to operate in all
atmospheric windows between 200 $\mu$m and 1 mm, and routinely at
350 and 450 $\mu$m throughout the year, has been the main objective of our site
testing activities since 2008, as reported in the
ARENA\footnote{Antarctic Research, a European Network for Astrophysics: 
http://arena.unice.fr} European conference proceedings.

\subsection{Prerequisites for a large submm telescope at Dome C} 

A study of the necessary pre-requirements for the future deployment of a large, submm
telescope has been carried out between 2007 and 2010 at
Concordia. Besides the atmospheric transmission in all
windows in the far-infrared and submillimetre parts of the spectrum,
complementary site testing has focused on the polar constraints and
their potential impact on instruments. This work includes a study of the removal and prevention of frost
formation, and a study of the temperature gradient in the ground layer.


%
%

\section{Ground-based measurement of atmospheric transmission}

The sky opacity at submillimetre wavelengths depends critically on the amount of
water vapour in the atmosphere. SUMMIT08 (SUbMilliMetre Tipper version 2008)
is an in-situ instrument measuring sky transparency at 200 $\mu$m. It
consists of a motorised mirror collecting atmospheric radiation at
different elevations.  This flux is then measured by a photometer behind a
200-$\mu$m filter. The previous prototype, SUMMIT, was installed
in 2000 at Concordia by the University of New South
Wales (UNSW) and performed measurements during the
Austral summer between December 2000 and January 2001
\citep{Calisse2004}. SUMMIT has then been refurbished in Saclay and
was installed at Concordia in February 2008 under
the responsability of CEA/Irfu in collaboration with UNSW \citep{Durand2008}.
The instrument then worked almost continuously between April 2008 and December 2008, and has been in
full operation since June 2009.

\subsection{The atmospheric model MOLIERE}  \label{moliere} 

MOLIERE (Microwave Observation and LIne Estimation and REtrieval) is a
forward and inversion atmospheric model \citep{Urban2004}, developed
for atmospheric science applications. The code calculates the
absorption of radiation in the mm- to mid-IR wavelength range
(equivalent from 0 to 10 THz in frequency) as a result of several
effects: (i) spectroscopic lines such as those of atmospheric water (H$_2$O), oxygen
(O$_2$), and ozone (O$_3$) absorb strongly at short
wavelengths, while (ii) collisions of H$_2$O with O$_2$ and
nitrogen (N$_2$) result in continuous absorption across all
wavelengths. For the line absorption, a radiative transfer model
including refraction and absorption by major and minor atmospheric
species is included (H$_2$O, O$_2$, O$_3$, NO$_2$, HNO$_3$, CO and other
lines up to 10 THz). Spectroscopic parameters are taken from
HITRAN\footnote{High Resolution Transmission molecular absorption
database} and the JPL\footnote{Jet Propulsion Laboratory} database.
The pseudo-continuum water vapour absorption and collisionally-induced 
dry absorption are included as a quadratic frequency term
according to \citet{Pardo2001} for f$<$1 THz and
\citet{Borysow1986} for f$>$1 THz.

This model has previously been used to calculate the atmospheric transmission
up to 2000 GHz ($\sim$150 $\mu$m) for a large number of astronomical
sites \citep{Schneider2009}. The results can be found on a dedicated
website\footnote{http://transmissioncurves.free.fr}.  For those studies,
approriate temperature and pressure profiles (for example Antarctic
profiles for Dome C) of the sites were used \citep{Schneider2009}. The colder Antarctic atmosphere preferentially populates the low-lying energy state of water, leading to a greater absorption for a given PWV than that of a non-polar site. This effect is included in the model.

\begin{figure}
\centering
\includegraphics[width=0.60\linewidth, angle=-90]{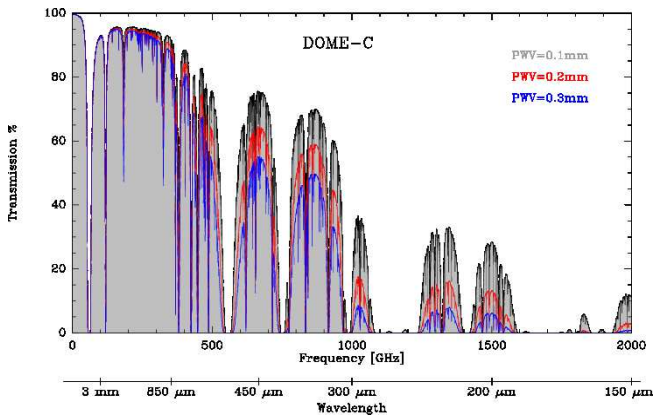}
\caption{
Atmospheric transmission at 0.1, 0.2, and 0.3mm PWV between 0--2 THz 
($\approx$3mm to $\approx$150 $\mu$m) at Dome C calculated with the atmospheric model MOLIERE.}
\label{moliere_fig}
\end  {figure}

\subsection{Measurements, data analysis and results}\label{data}

\begin{figure*}
\centering
\includegraphics[width=\linewidth]{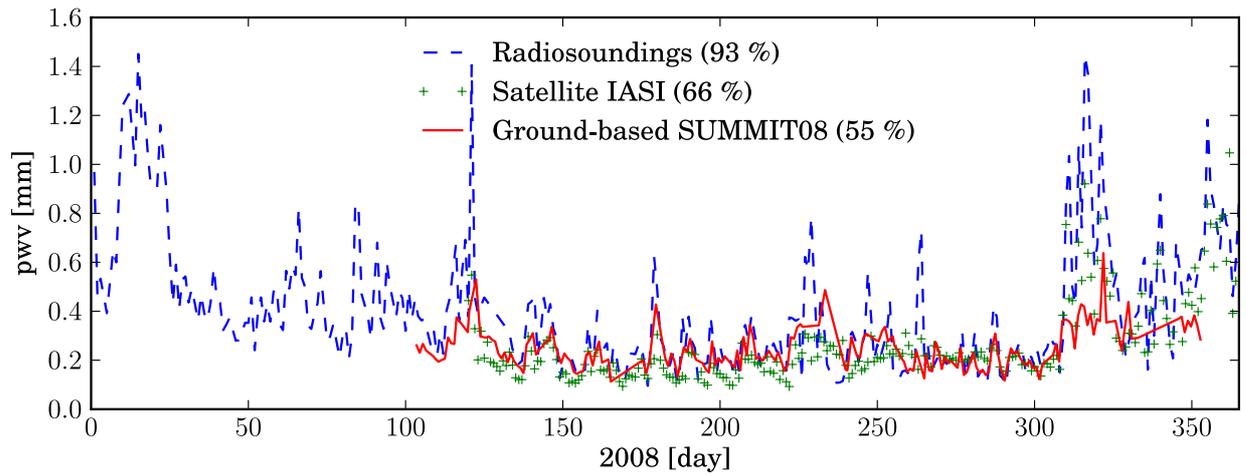}
\caption{Precipitable Water Vapour (PWV) in [mm] in 2008 at the Concordia Station measured in-situ by SUMMIT08 (red line), by radiosoundings (blue line) and from 
space by IASI (green line). Percentages indicate the fraction of time that each of the instruments worked during the year.}
\label{2008}
\end{figure*}

\begin{figure*}
\centering
\includegraphics[width=\linewidth]{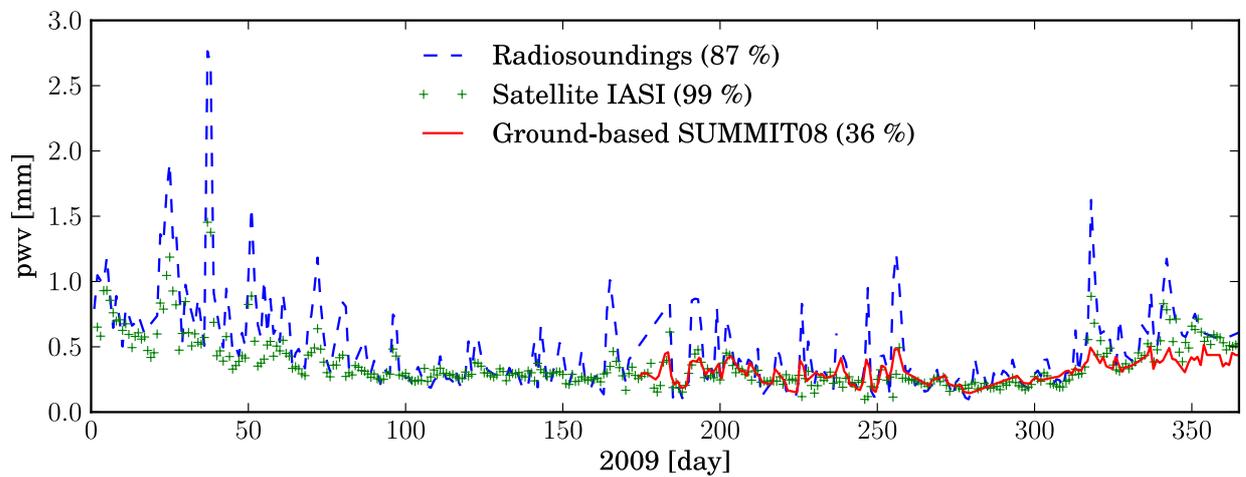}
\caption{PWV in 2009 at the Concordia Station 
measured in-situ by SUMMIT08 (red line), by radiosoundings (blue line) and from space by IASI (green line).}
\label{2009}
\end{figure*}

\begin{figure*}
\centering
\includegraphics[width=\linewidth]{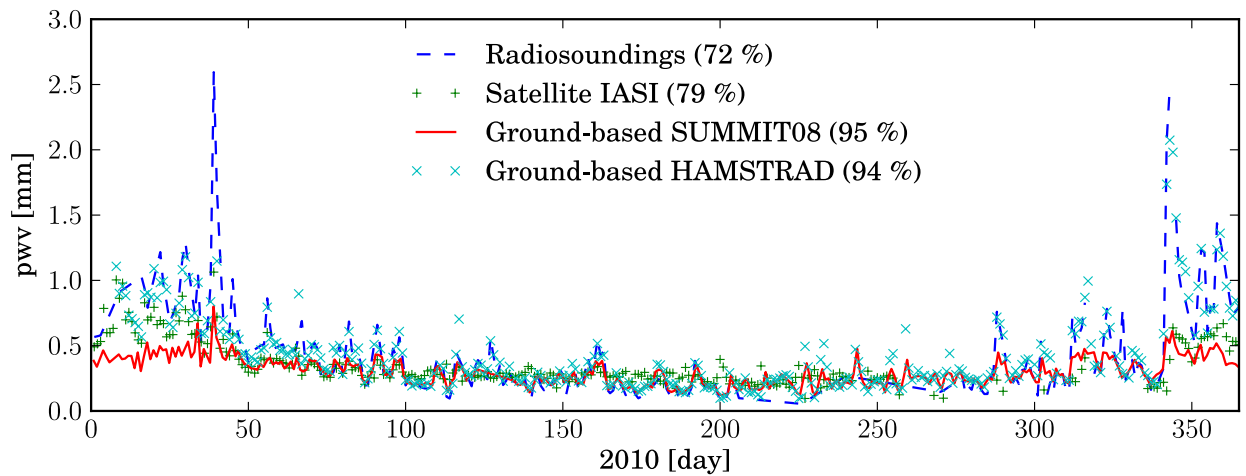}
\caption{PWV in 2010 at the Concordia Station 
measured in-situ by SUMMIT08 (red line), by radiosoundings (blue line), by HAMSTRAD (turquoise line) and from space by IASI(green line).}
\label{2010}
\end{figure*}

The atmospheric emission at 200 $\mu$m $S(X)$ is a function of zenith opacity $\tau_{200}$, airmass
$X$ and atmospheric temperature $T_{atm}$ (see Eq. \ref{satm}). $G$ and $T_{ref}$ are the gain and the offset 
of the detector determined by an internal ``hot-cold'' calibration. $a_{eff}$ is the transmission of an effective grey body at temperature $T_{eff}$
which takes in account all the noise contributions \citep[e.g. the instrument window, see][]{Calisse2004}; these parameters were determined by an external calibration  performed 
during the two summer campaigns. The two unknown variables  $\tau_{200}$ and $T_{atm}$ can be extracted from Eq. \ref{satm} by a non linear regression 
of $S$ as function of the airmass.

\begin{equation}\label{satm}
  S(X)=-G(a_{eff}(1-e^{-X\tau_{200}})T_{atm}+(1-a_{eff})T_{eff}+T_{ref})
\end{equation}

The 200-$\mu$m filter has a bandwidth (FWHM) of 9 $\mu$m. 
The measured zenith transmission is an average over the bandwidth of the
filter.
Using the atmospheric model MOLIERE (Sect. \ref{moliere}), we
characterize the atmospheric window at a given PWV, i.e. the
transmission of the atmosphere as a function of the wavelength.  
For example, in Fig. \ref{moliere_fig}, the average value of the transmission over the bandwdith of the atmospheric 
window around 200 $\mu$m is less than the transmission at the peak. 
We can convolve the transmission spectra of
the filter with that of the atmospheric window and thereby evaluate how much the peak transmission is under-estimated. 
On average, a relative under-estimation of 25\% was calculated for the zenith transmission 
(i.e. an average transmission of 10\% implies a peak transmission of 12.5\%).\\

MOLIERE is used to relate the opacity at any wavelength to the
atmospheric PWV. For the 200-$\mu$m window, for example, we extract
from the transmission curves (Fig. \ref{moliere_fig}) the 200-$\mu$m
transmission for a given PWV. Then, we convert the transmission into
an opacity and perform a linear regression of the PWV as a function of
this opacity.  The relation is given by equation \ref{pwv} with a
correlation coefficient of 0.9999.
\begin{equation}\label{pwv}
PWV=0.13{\times}\tau_{200}-0.06
\end{equation}

SUMMIT08  works on a two-hour cycle, performing a measurement at
airmass $X$=1 and 2 as well as two full skydips with ten pointing 
angles from horizon to zenith. The
transmission at 350 $\mu$m and the PWV were deduced using MOLIERE,
with the method described above.

The points in Figs. \ref{2008},
\ref{2009}, and \ref{2010} are average values per day, for 2008, 2009,
and 2010 respectively. Quartiles of transmission at zenith for 200 and
350 $\mu$m (Trans$_{200/350}=\exp(-\tau_{200/350})$), and PWVs are
indicated in Table. \ref{quartile}. Quartiles represent the
distribution function of PWV content, e.g. there is less than 0.19 mm
of PWV at Dome C, 25\% of the time that SUMMIT08 was working in 2008. 

The quartiles indicate that there was a degradation of the transmission between
2008 and 2009 as the amount of PWV increased significantly in one year.

\begin{table}[!ht]
\caption{\label{quartile} Quartiles of SUMMIT08 measurements in 2008, 2009, 2010 and cumulated on the three years.
PWV and transmission at 350 $\mu m$ are derived using the MOLIERE model.}
\centering
\begin{tabular}{l|l|l|l}
     
 Time \% 2008 & PWV (mm)  & $Trans_{200} $ & $Trans_{350}$ \\
\hline
25 \% & 0.19 & 0.15         & 0.60         \\
50 \% & 0.22 & 0.11         & 0.56         \\
75 \% & 0.29 & 0.07         & 0.50         \\
\hline
\hline
Time \% 2009 & PWV (mm)  & $Trans_{200} $ & $Trans_{350}$ \\
\hline
25 \% &  0.25 & 0.09          & 0.54          \\
50 \% &  0.32 & 0.05          & 0.48          \\
75 \% &  0.39 & 0.03          & 0.42          \\
\hline
\hline
Time \% 2010 & PWV (mm)  & $Trans_{200} $ & $Trans_{350}$ \\
\hline
25 \% &  0.22 & 0.12          & 0.57          \\
50 \% &  0.28 & 0.07          & 0.51          \\
75 \% &  0.38 & 0.03          & 0.43          \\
\hline
\hline
Time \% cumulated & PWV (mm)  & $Trans_{200} $ & $Trans_{350}$ \\
\hline
25 \% &  0.21 & 0.12          & 0.58          \\
50 \% &  0.27 & 0.08          & 0.52          \\
75 \% &  0.35 & 0.04          & 0.45          \\
\end{tabular}
\end{table}

\begin{figure}
\centering
\resizebox{0.8\hsize}{!}{\includegraphics{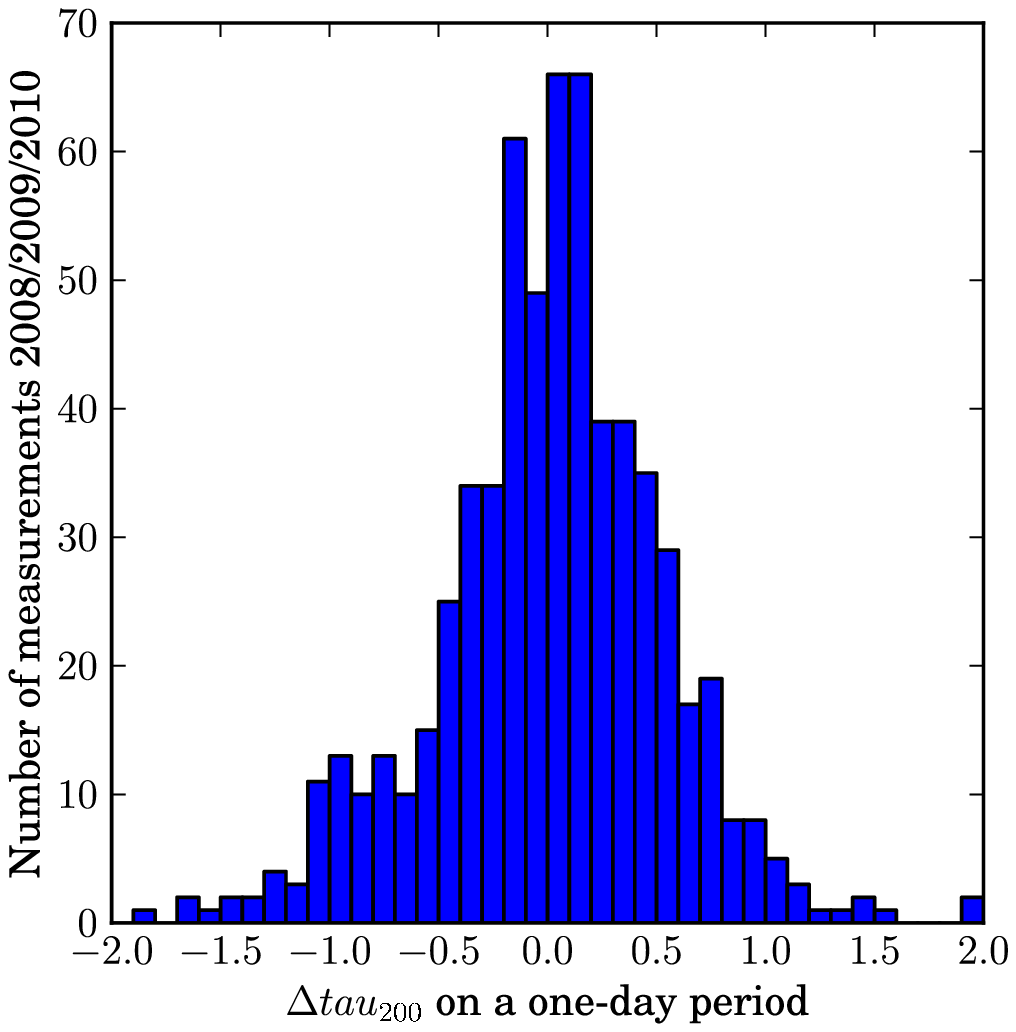}}
\caption{Distribution of 200-$\mu$m opacity variations on a one-day statistic 
for 2008/2009/2010}
\label{distribution}
\end  {figure}

Opacity variations based on a one-day avearges are plotted in
Fig. \ref{distribution}. The standard deviation of the opacity
distribution is 0.45, which corresponds to variations of only 37\% in
transmission. Atmospheric conditions are
very stable during winter in Antarctica, where the polar night avoids
the strong diurnal variations of water vapour induced by the solar radiation
in non-polar regions.

\section{Comparisons between instruments}

\subsection{Satellite measurements : IASI} \label{iasi}

\begin{figure}
\centering
\resizebox{0.8\hsize}{!}{\includegraphics{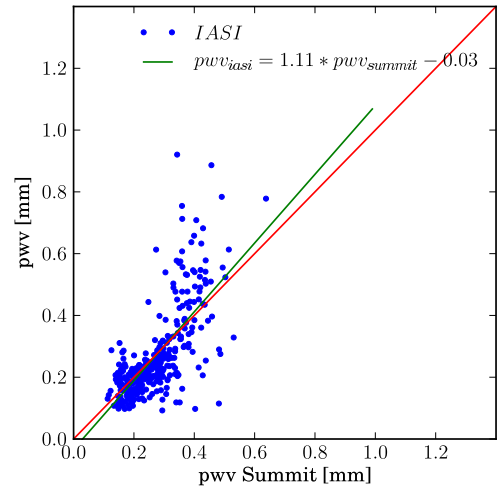}}
\caption{Correlation between in-situ measurement (SUMMIT08) and satellite 
measurments(IASI) between 2008 and 2010}
\label{corr_iasi}
\end  {figure}

IASI (Infrared Atmospheric Sounding Interferometer) is an atmospheric
interferometer working in the infrared, launched in 2006 on the
 METOP-A satellite \citep{Phulpin2007,Pougatchev2008}. The data are
available at the website of the Centre for Atmospheric Chemistry
Products and Services\footnote{http://ether.ipsl.jussieu.fr}. IASI was
developed by the French Space Agency CNES\footnote{Centre National
d'Etudes Spatiales} in collaboration with EUMETSAT\footnote{European
organisation for meteorological satellites
http://www.eumetsat.int}. The satellite is in a polar orbit such that
each point on Earth is seen at least once a day by the detector. IASI is a
Fourier Transform Spectrometer (FTS) working between 3.7 and 15.5
$\mu$m. It is associated with an infrared imager, operating between
10.3 and 12.5 $\mu$m. Each pixel of the instrument corresponds to a spatial extent of
 12 km at nadir, and vertical profiles of humidity at ninety altitude levels
are retrieved with a typically 10\% accuracy. The amount of precipitable water 
vapour is given by the integral of these
vertical profiles. We took all the measurements in a zone of 110 km$^2$
around Concordia. The results for 2008, 2009, and 2010 are plotted
together with the SUMMIT08 data in Fig. \ref{2008}, \ref{2009}, and
\ref{2010} together with the correlation of the data in Fig. \ref{corr_iasi}.
There is a small negative bias for SUMMIT08
(SUMMIT08 values are lower than the ones for IASI) at low PWV (less than
0.4 mm), and a significant negative bias at high PWV (more than 0.4 mm). This bias occurs 
during the summer period, when the PWV is high. An analysis of the skydips shows that
the fluxes received at different elevations are highly variable during the summer and do not fit 
the exponential profile given by equation \ref{satm}. 
Therefore we conclude that there are atmospheric temperature or PWV variations at the scale of Summit field
of view that pollute the analysis and prevent us to extract an accurate value for the PWV during summer periods.

\subsection{In-situ measurements: Radiosoundings \& HAMSTRAD}

\begin{figure}
\centering
\resizebox{0.8\hsize}{!}{\includegraphics{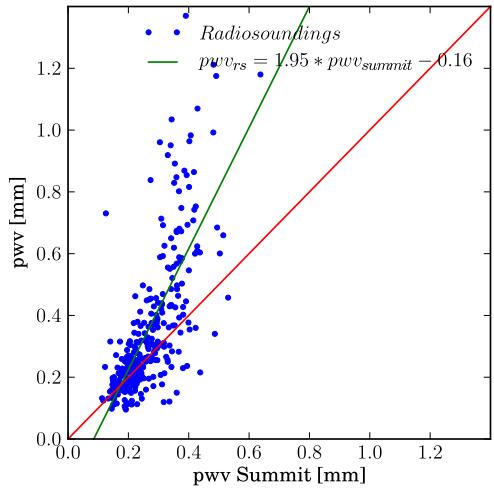}}
\caption{Correlation between in-situ measurements SUMMIT08 and radiosoundings 
 between 2008 and 2010}
\label{corr_rs}
\end  {figure}

An atmospheric radiosounding station was installed at Concordia
Station in 2005.  Radiosonde balloons are launched twice a day, and
reach an altitude of 25-30 km.  The vertical profile of the atmospheric
pressure, temperature and relative humidity is measured during
the ascent. PWV is calculated by integrating this vertical profile.
The data and additional information were obtained from IPEV/PNRA Project ''Routine
Meteorological Observation at Station Concordia
\footnote{http://www.climantartide.it}'' \citep[see][]{Tomasi2006}.  The results from these measurements are plotted in
Fig. \ref{2008}, \ref{2009}, and \ref{2010} and the correlation of the
data with SUMMIT08 data in Fig. \ref{corr_rs}.

\begin{figure}
\centering
\resizebox{0.8\hsize}{!}{\includegraphics{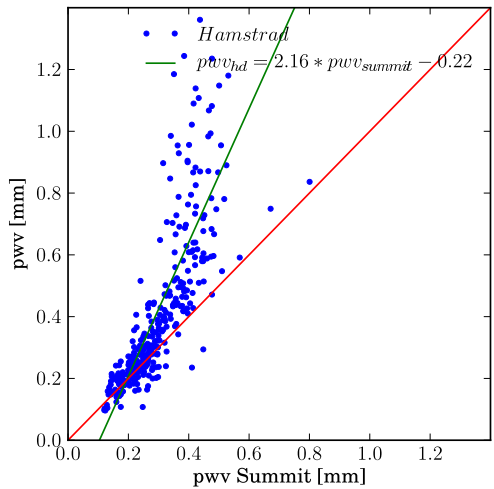}}
\caption{Correlation between in-situ measurements SUMMIT08 and HAMSTRAD in 2010}
\label{corr_hd}
\end  {figure}

HAMSTRAD is a microwave instrument operating at 60 and 183 GHz,
measuring temperature and water vapour profiles between 0 and 10 km with
a time resolution of 7 minutes \citep{Ricaud2010}. The instrument has been
fully operating since the beginning of 2010. PWV derived from the
vertical profile is plotted in Fig. \ref{2010}, and the correlation of
these data with the SUMMIT08 data in Fig. \ref{corr_hd}.  The SUMMIT08
measurements and radiosonde or HAMSTRAD data are well correlated
at low PWV (less than 0.4 mm), but less so at high PWV (more
than 0.4 mm), probably due to temperature or PWV spatial variations during summer skydips of SUMMIT08.

\subsection{Instrument and site comparison}

\begin{figure}
\centering
\includegraphics[width=4.4cm]{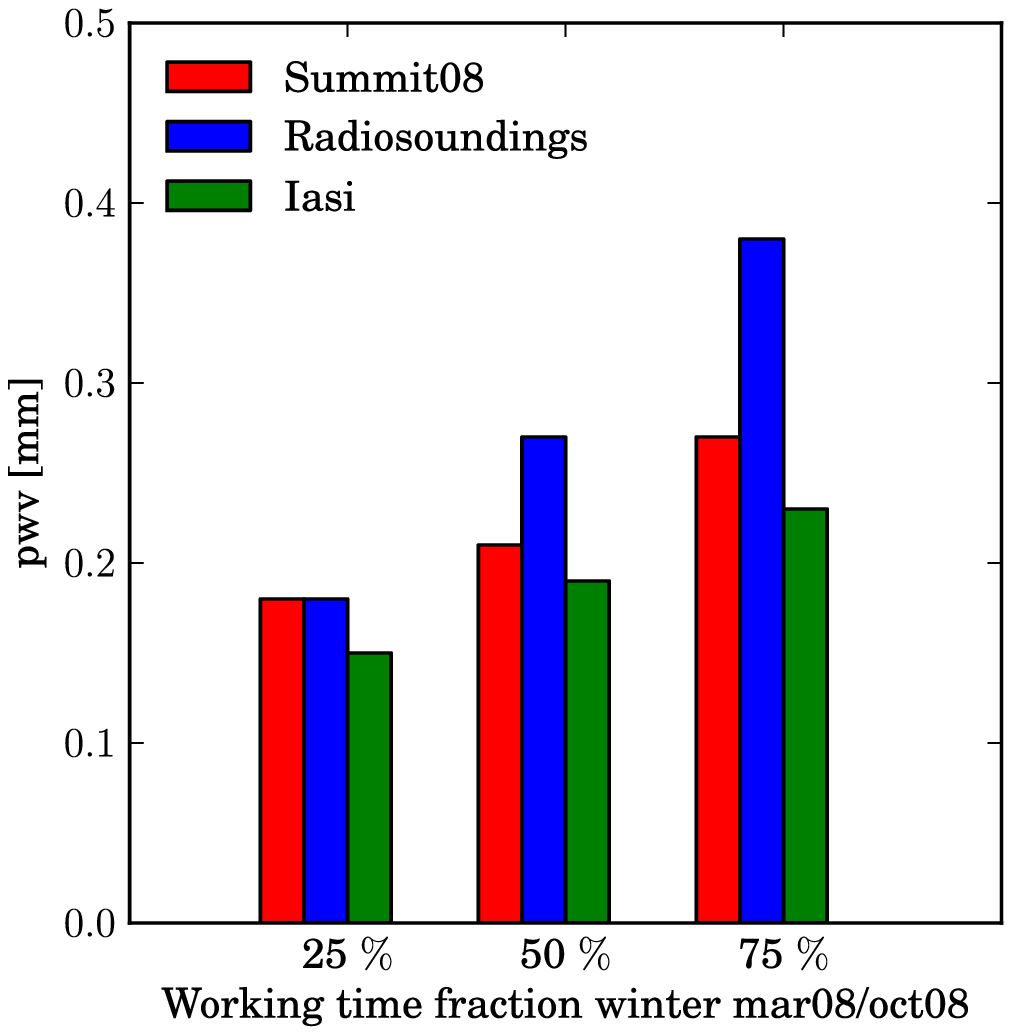}
\includegraphics[width=4.4cm]{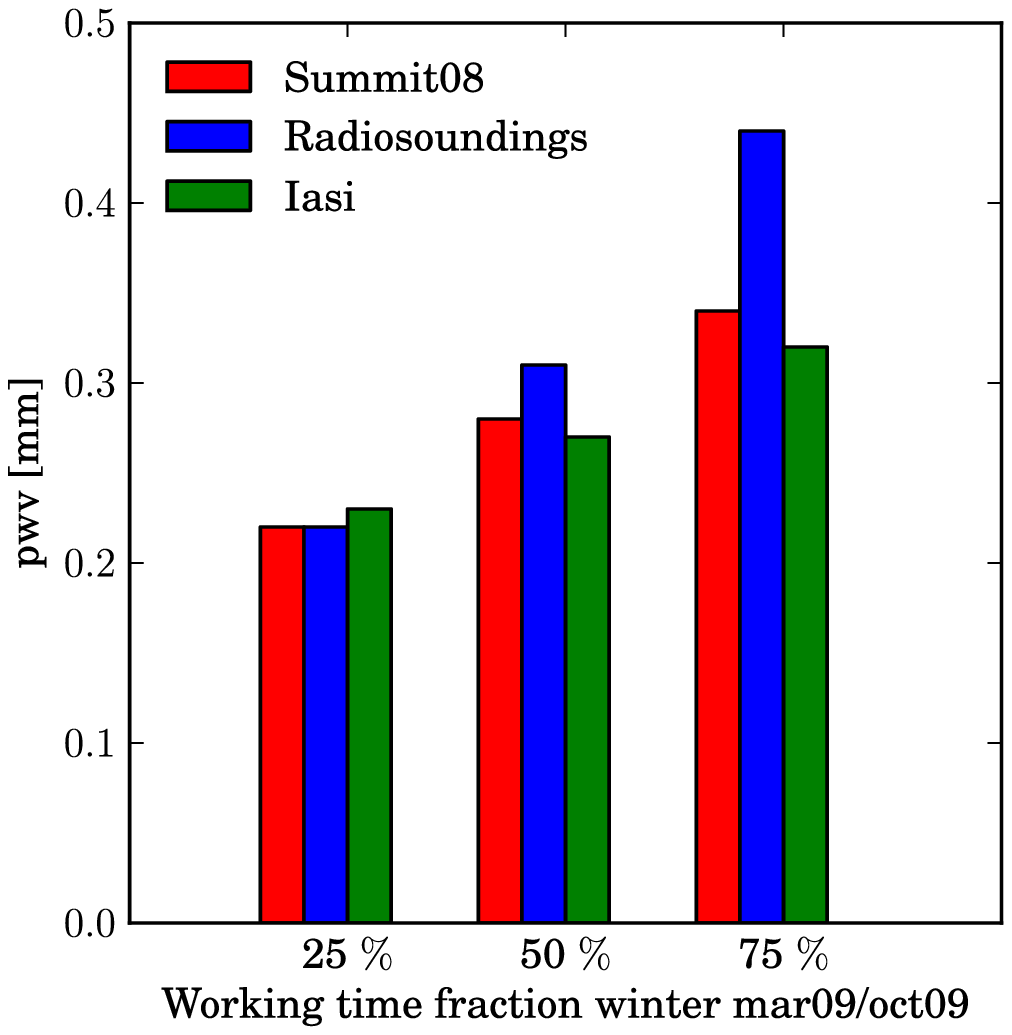}
\includegraphics[width=4.4cm]{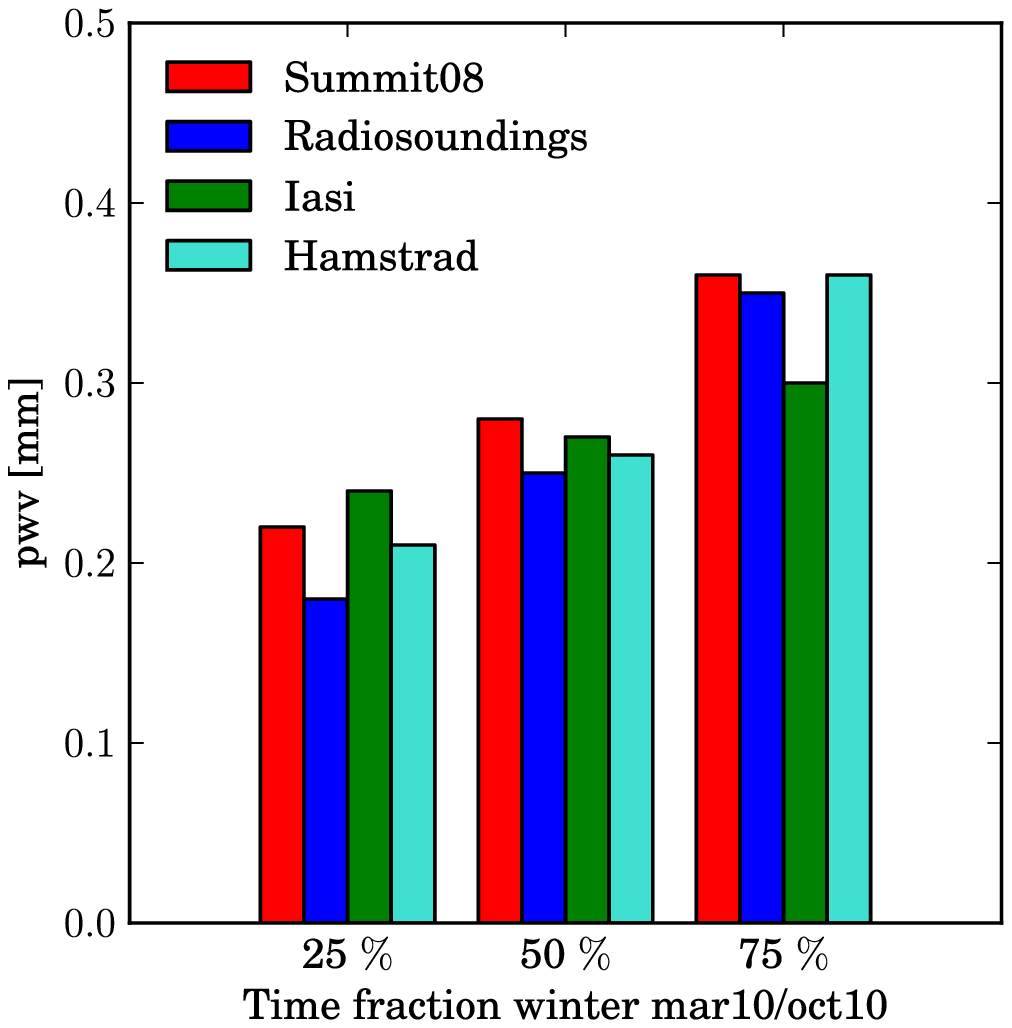}
\caption{Winter quartile during the working time of the instruments between 
March 2008 and October 2010.}
\label{winter}
\end  {figure}

\begin{figure}
\centering
\includegraphics[width=4.4cm]{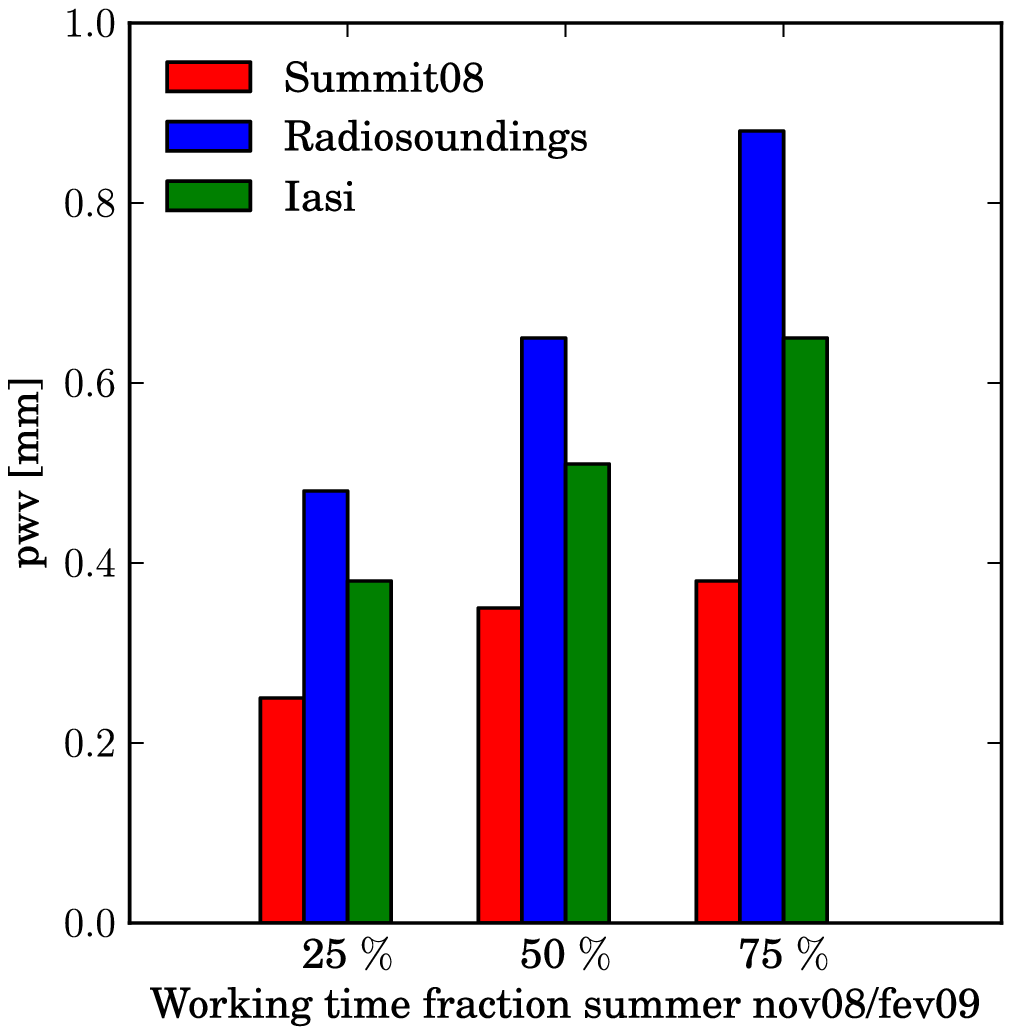}
\includegraphics[width=4.4cm]{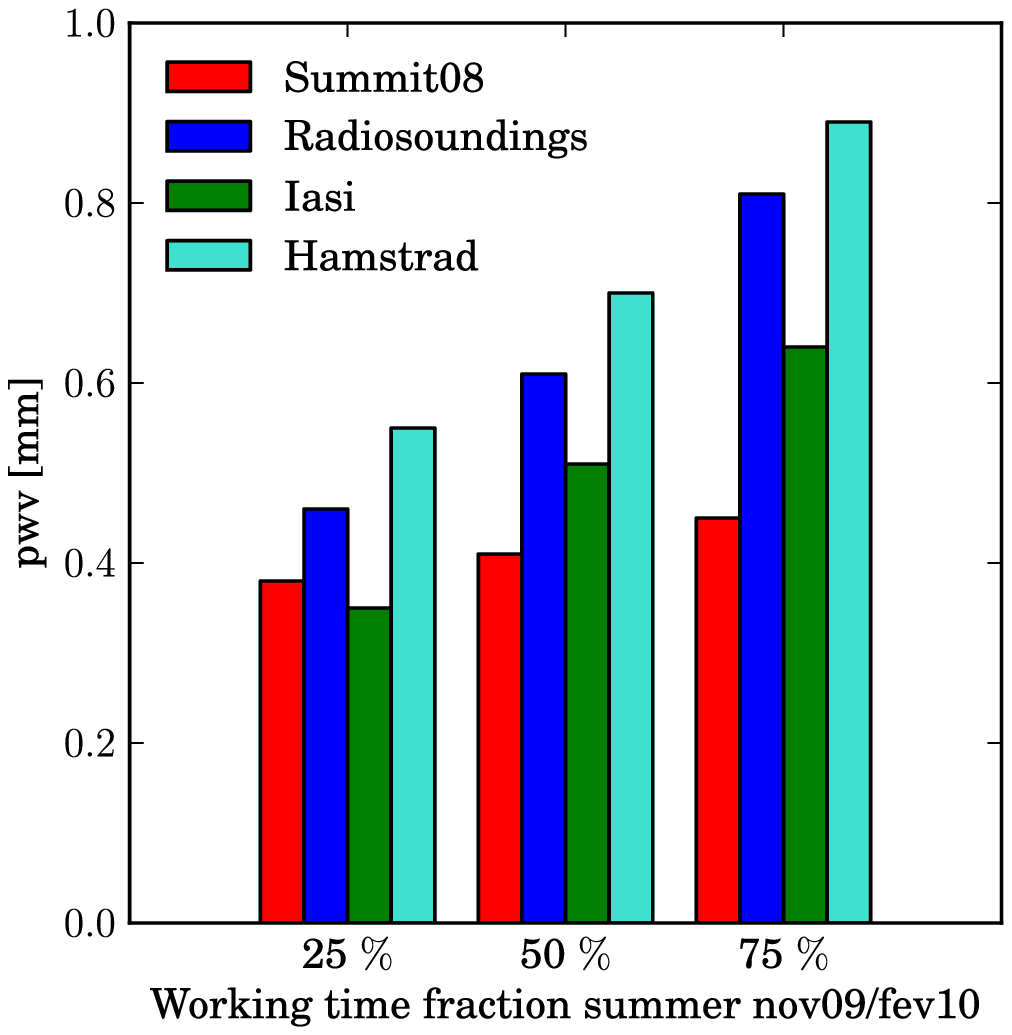}
\includegraphics[width=4.4cm]{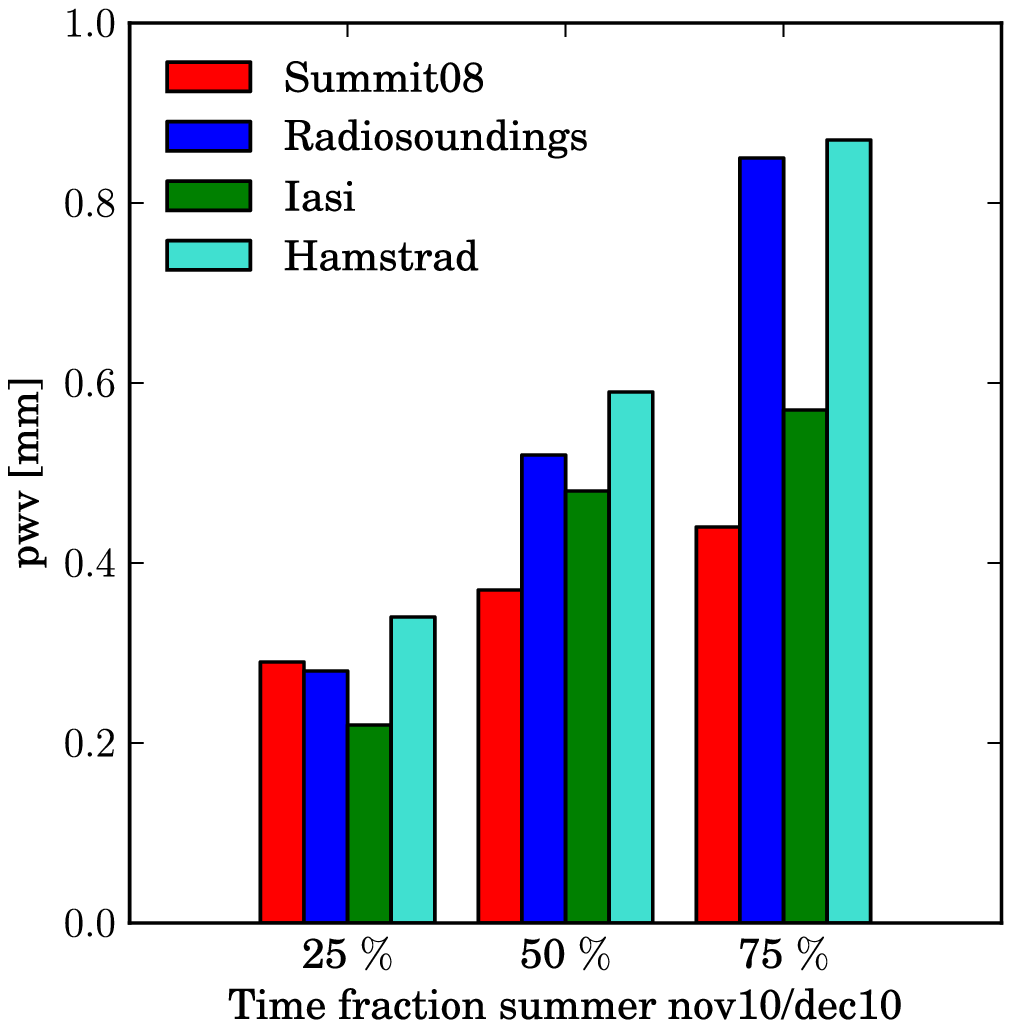}
\caption{Summer quartile during the working time of the instruments between 
March 2008 and December 2010.}
\label{summer}
\end  {figure}

\begin{figure}
\centering
\includegraphics[width=4.4cm]{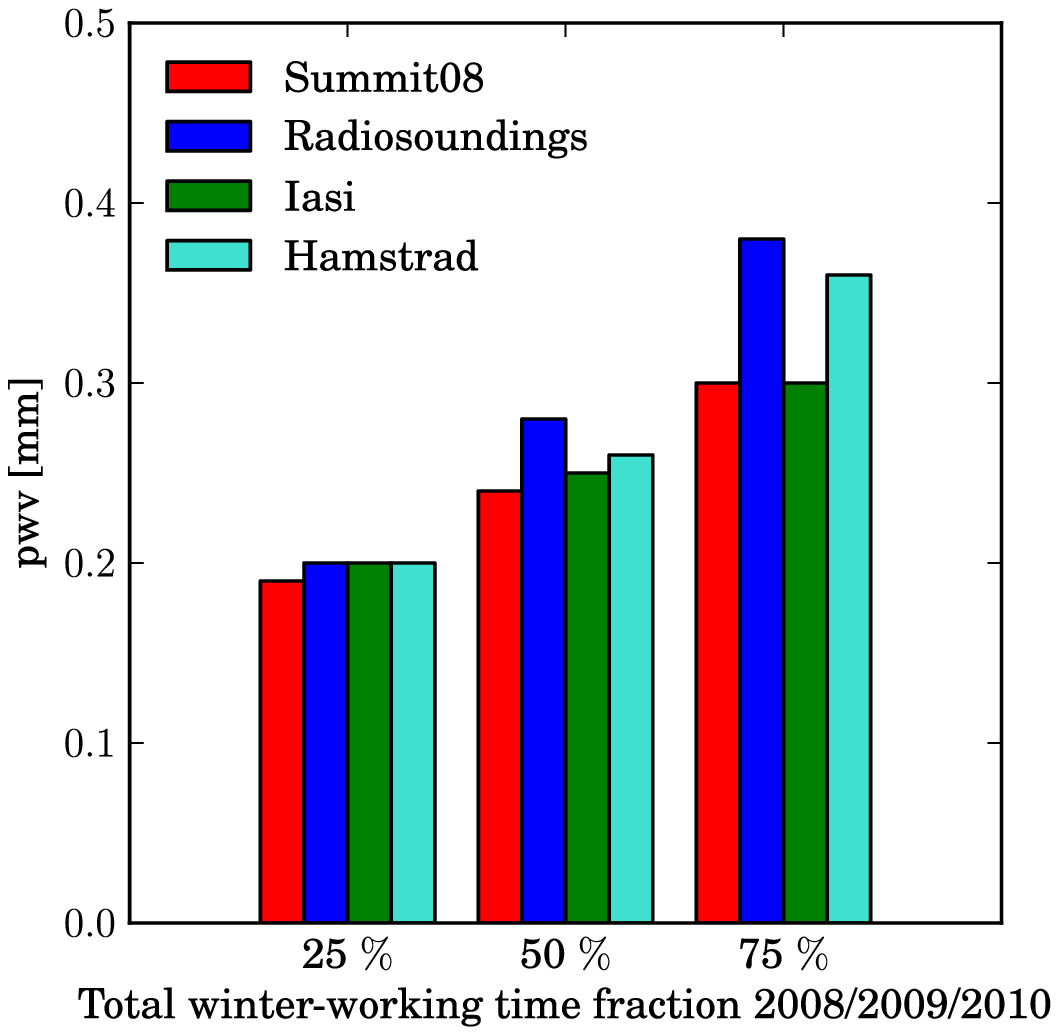}
\includegraphics[width=4.4cm]{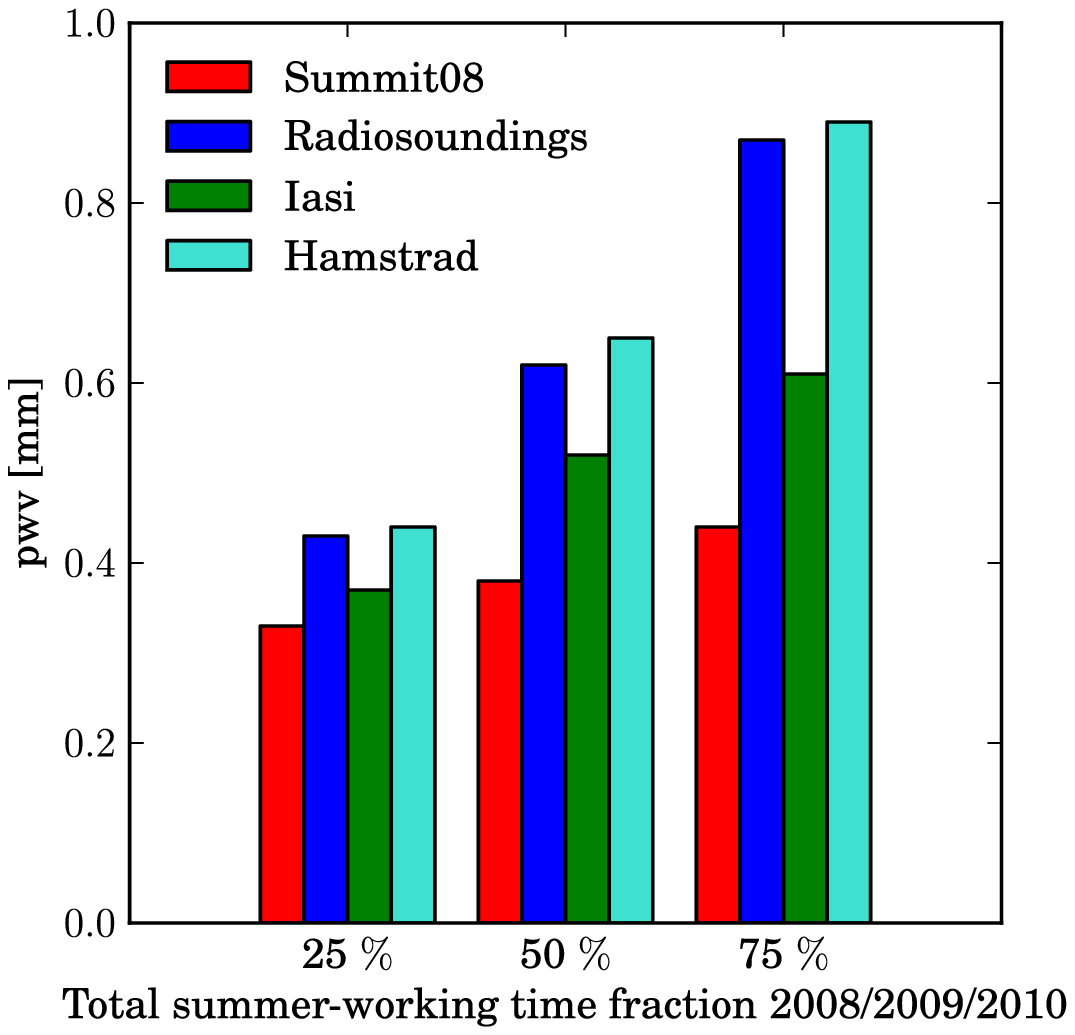}
\caption{Total quartile per summer/winter periods during the working time of the 
instruments between March 2008 and December 2010.}
\label{total}
\end  {figure}

Low PWV periods mainly occur during the austral winter (from March to October) 
and high PWV periods are concentrated during austral summer periods (from November to
February). In
contrast to, for example, the Chilean sites where strong day-to-day
variations are found, there is a very long and stable period of high
transparency at submillimetre wavelengths at Dome C.  This will be the subject an upcoming paper, in which we
study a large number of locations using satellite data to
characterize the quality of the sites in an unbiased way. The stable
periods at Dome C are identified as \textit{winter periods} in the
quartiles (Fig. \ref{winter}). \textit{Summer period} quartiles are
plotted in Fig. \ref{summer}, and total quartiles per period in
Fig. \ref{total}. The time fractions used for these quartiles are
based on the time the instruments were in operation. The quartiles of
SUMMIT08 are slightly biased because it was not functional all the
time during these periods (between 36\% and 95\%), while the
satellite and radiosounding quartiles are less biased (IASI was operational
between 85\% and 100\% of the time during these periods).

The SUMMIT08 campaign allows validation of the satellite
measurements. There is a very good
correlation during winter periods between IASI, SUMMIT08, HAMSTRAD,
and radiosonde data. However the correlation is not so good during summer periods,
or more generally when PWV is high (more than 0.4 mm). HAMSTRAD is well correlated with
the radiosonde data but there is a dry bias for IASI data. This bias was previously identified
by \citet{Ricaud2010} and could be attributed to the influence of the surface emission parameter
over Antarctica. In order to
further improve the effectiveness of satellites for measuring temperature
and humidity over the White Continent, a consortium
\citep[see][]{Rabier2010} was created to collect ground-based data
for comparison; in particular to understand why measurements differ at high PWV. 

Nevertheless, 
for astronomical observations, we are only interested in PWV values below about 0.3 mm 
(50\% transmission at 350 $\mu$m wavelength), and the disparity between measurements at high PWV 
disappears for the median and 25\% winter time fraction (Fig. \ref{winter} and Fig. \ref{total}).

%
%

\section{Polar Constraints}

\subsection{Frost Formation}

The GIVRE experiment was designed in 2006 to study frost formation
\citep{Durand2008}. It consists of a series of twelve probes of
different shapes and colors: five black and two bright cylinders, two
black and two bright disks, and an aluminium plate. The probes are divided
in two groups, which were placed at different altitudes on a
fifty meter tower. They can all be warmed by heaters except for the plate,
which instead can have dry air (from near the ground) blown over it. In 2007,
building upon the results from the first year of tests, an antifrost system was
installed on the COCHISE telescope \citep{Sabbatini2009} at
Concordia. Three different techniques are used : heaters at the rear of the
mirror, an infrared lamp, and an air blowing system.

\begin{table}[!ht]
\caption{\label{pressure} Water vapour partial pressure for saturated air as a function of atmospheric temperature.}

\centering
\begin{tabular}{c|c}
Temperature ($^\circ$C) & Partial Pressure Of Water Vapour (Pa) \\
\hline
-40 & 12.84 \\
-50 &  3.94 \\
-60 &  1.08 \\
-70 &  0.26 \\
\end{tabular}
\end{table}

During the summer season, the mean temperature at Concordia is of the
order of -30$^\circ$C and no frost appears on instruments during this
period. However during winter, the temperature can fall to
-70$^\circ$C and the partial pressure of water vapour is two orders of magnitude lower (see
Table \ref{pressure}). At -70$^\circ$C, a partial pressure of water vapour of 0.26 Pa
corresponds to an absolute humidity of 1 mg of water per m$^3$; i.e., a hundred times smaller than a drop of water. In such conditions, however, the
relative humidity is very high, nearly 100 \%. If the temperature
decreases by 5$^\circ$C, the relative humidity will increase by almost a factor of two and can become much higher than 100 \%,
leading to ice nucleation and frost formation. Frost can also appear
due to snow precipitation on surfaces.

Three systems can be used against frost 
formation and snow deposit:
\begin{itemize}
\item Warming up the probes by three or four degrees is enough to prevent frost. 
It is also the most efficient way to prevent frost formation on the
mirror of a telescope.
\item Taking dry air at snow surface level and blowing it over the probes is 
effective; it is also a useful method because it does not require
warm optics (which can introduce undesirable temperature gradients).
\item Explosive air blowing can be used to physically remove the snow deposit. 
\end{itemize}

\begin{figure*}
\centering
\includegraphics[width=\linewidth]{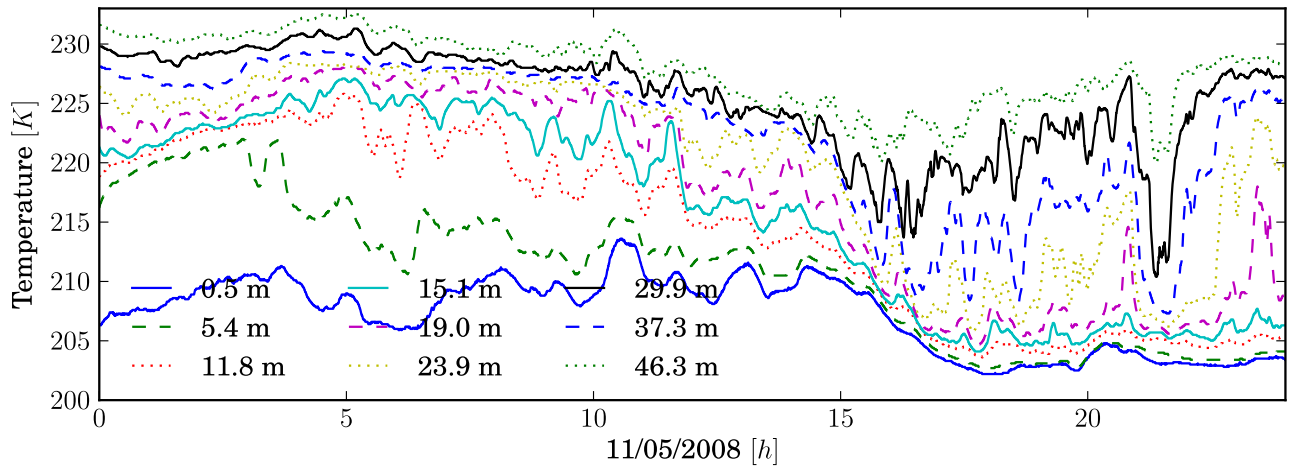}
\caption{Typical temperature variation at altitudes between 0.5 m and 46.3 m during a winter day. 
Each curve is the temperature in Kelvin measured at a given altitude on the 50m tower}
\label{temp}
\end  {figure*}

Frost mainly forms as a result of the cooling of a surface radiating towards
the sky. Close to these surfaces, the maximum amount of water vapour that the atmosphere can hold is
lower and frost appears. Warming up these surfaces by three or four
degrees is sufficient to compensate for the heat loss due to radiation. Blowing
dry air taken from close to the ground is an also effective way to dehumidify the local
environment.

\subsection{Temperature gradient in the boundary layer}

The large temperature gradient in the first thirty meters above the snow limits the
choice of materials that can be used to build large instruments in a polar
environment. The MAST experiment, installed in 2007 at Concordia,
consists of sixteen temperature probes placed on a fifty-meter tower
\citep{Durand2008} with a logarithmically increasing spacing bewteen the probes 
The probes are Pt1000 sensors sandwiched between two bright plates to protect them from solar 
radiation. GIVRE and MAST can be controlled from the station, and data
are automatically sent daily to the station and thence to Europe via the
station's communication system. \\

\begin{figure}
\centering
\resizebox{0.8\hsize}{!}{\includegraphics{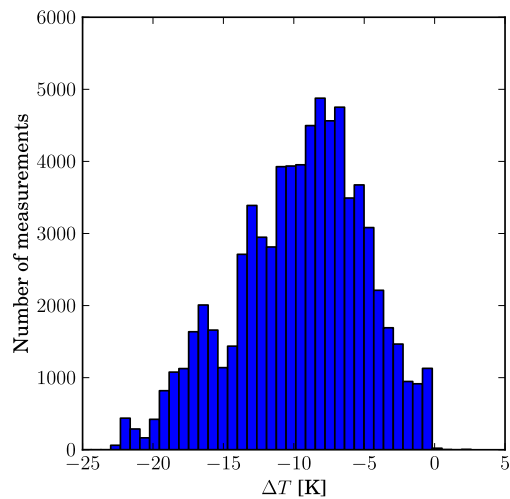}}
\caption{Distribution of temperature gradient between 0.5 m and 46.3 m 
between 01/2010 and 04/2010.}
\label{mast}
\end{figure}

Close to the ground, the atmosphere is cooled by contact with the snow
surface. The temperature increases with altitude up to a point of
inversion, above which the atmosphere has the usual tropospheric profile, decreasing by 6.5 $^\circ$C per kilometre. The atmosphere
between the ground and the point of inversion is called the boundary
layer. The thickness of this layer can be very low, leading to very
good astronomical seeing in the visible
\citep[see][]{Fossat2005}. However this thickness is variable and is
responsible for strong (vertical and temporal) temperature gradients close to the
ground. These gradients can be extreme, as shown in
Fig. \ref{temp} and \ref{mast}. The typical variations during a winter day can result
in a vertical gradient of 30$^\circ$C over 50 meters, and variations of 15$^\circ$C
in one hour have been recorded. However, the engineering issues this creates can be readily 
overcome by using materials with low thermal expansion.

%
%

\section{Discussion and conclusions}

The necessary requirements for the future deployment of a large (10--25 m) 
telescope have been explored between 2007 and  2010.
The atmospheric opacity at 200 $\mu$m was monitored for three years,
making this the first extended study at this wavelength and providing the first
multi-year statistics.
Based on these atmospheric opacity measurements and the derived PWV, we
can conclude that the submm/THz atmospheric windows open during a
relatively large and stable fraction of time, and particularly for 
all windows down to 350 $\mu$m. The 200-$\mu$m window
opens for somewhat less than 25\% of the time (PWV$<0.2$ mm). These findings
are based upon a $\sim$3 year study in which we noticed significant
variations from one year to another.

Comparisons between Dome C and other sites (e.g. Dome A, Chajnantor)
should be made over the same period of time to be directly comparable.  In addition,
the issue of calibration \citep[window transmission, filter bandwdith,. etc see][]{Radford2009} 
usually complicates comparisons between sites where measurements are made with different instruments. 
However, although non-polar sites can offer
very good conditions for submillimetre observations (e.g. Chajnantor
\citet{Radford2008}), Antarctica provides not only better
transmission, but also extremely good stability that is invaluable for time-series studies at submillimetre wavelengths.

In addition to the measurement of the atmospheric opacity, a
knowledge of the polar constraints is an equally necessary
prerequisite for operating in the harsh Antarctic plateau environment. The thermal
gradient in the boundary layer and the problem of ice formation have been extensively
monitored. Techniques for preventing or removing ice have also been
implemented on site. Solutions for preventing and removing icing and
frost proved to be useful on probes and on the surface of a 2.5-m
diameter telescope, which could scale-up for installations on larger
and more complex instruments. Another polar constraint of importance
is the thermal gradient that was measured in a 46-m high layer above the
ground. \citet{Radford2009} pointed out the possible lack of surface
accuracy of mirrors due to thermal gradients arising from both the vertical
gradient and iced surface of the telescope. Tests of image quality
while removing frost will be done on the IRAIT telescope
\citep{Tosti2006} that is currently deployed at Dome C. Further
studies of the polar constraints based on the thermal gradient and
variation will be discussed in technical reports. For the scope of
this paper, we may conclude that any instrument in the boundary layer
should be built with materials whose thermal expansion coefficient is low
at temperatures around -70$^\circ$C. In addition, materials must be able to be cooled rapidly to very low temperature without
internal stresses being created. 

In conclusion, Dome C is currently one of the best
accessible sites on Earth in terms of transmission in the
FIR/submm. The 200-$\mu$m window opens at a level of 10-15\% for slightly less than 25\%
of the time. Further analysis indicates that observations at 350 and
450 $\mu$m would be possible all year round. These very good
astronomical conditions in terms of atmospheric transmission and
stability offer a unique opportunity for high angular resolution and time series observations at
Dome C with medium size instruments.   

%
%

\begin{acknowledgements}

We acknowledge the strong support of the French and Italian polar 
institutes IPEV and PNRA that make this research project possible. We also 
acknowledge the European network ARENA. The original SUMMIT instrument was designed and built by Jeff Peterson and Simon Radford.

\end{acknowledgements}

%
%

\bibliographystyle{aa}
\bibliography     {bibliography.bib}


\end {document}